\documentclass[twocolumn]{aastex61}

\received{}
\revised{}
\accepted{}
\submitjournal{ApJ}

\shorttitle{Constraining BH mergers in AGN disks}
\shortauthors{McKernan et al.}

\begin{document}

\title{Constraining stellar mass black hole mergers in AGN disks detectable with LIGO}

\correspondingauthor{Barry McKernan}

\author{Barry McKernan}
\affil{Dept. of Science, CUNY-BMCC, 199 Chambers St., New York NY 10007}
\affil{Dept. of Astrophysics, American Museum of Natural History, Central Park West, New York, NY 10028}

\author{K.E.Saavik Ford}
\affil{Dept. of Science, CUNY-BMCC, 199 Chambers St., New York NY 10007}
\affil{Dept. of Astrophysics, American Museum of Natural History, Central Park West, New York, NY 10028}

\author{J.Bellovary}
\affiliation{Dept. of Physics, CUNY-QCC, Bayside, New York NY 11364}
\affil{Dept. of Astrophysics, American Museum of Natural History, Central Park West, New York, NY 10028}

\author{N.W.C. Leigh}
\affil{Dept. of Astrophysics, American Museum of Natural History, Central Park West, New York, NY 10028}

\author{Z. Haiman}
\affil{Columbia Astrophysics Laboratory, Columbia University, New York, NY 10027}

\author{B. Kocsis}
\affil{Institute of Physics, E\"{o}tv\"{o}s University, Budapest 1117, Hungary}

\author{W. Lyra}
\affil{Dept. of Astrophysics, American Museum of Natural History, Central Park West, New York, NY 10028}

\author{M.-M. Mac Low}
\affil{Dept. of Astrophysics, American Museum of Natural History, Central Park West, New York, NY 10028}

\author{B. Metzger}
\affil{Columbia Astrophysics Laboratory, Columbia University, New York, NY 10027}

\author{M.O'Dowd}
\affiliation{Dept. of Physics, CUNY-Lehman, New York NY 10468}
\affil{Dept. of Astrophysics, American Museum of Natural History, Central Park West, New York, NY 10028}

\author{S. Endlich}
\affil{Stanford Institute for Theoretical Physics, Stanford University, CA 94306}

\author{D.J.Rosen}
\affil{Dept. of Astrophysics, American Museum of Natural History, Central Park West, New York, NY 10028}



\begin{abstract}
Black hole mergers detectable with LIGO can occur in active galactic nucleus (AGN) disks. Here we parameterize the merger rates, the mass spectrum and the spin spectrum of black holes (BH) in AGN disks. The predicted merger rate spans $\sim 10^{-4}-10^{4} \rm{Gpc}^{-1} \rm{yr}^{-1}$, so upper limits from LIGO ($<212\rm{Gpc}^{-1}\rm{yr}^{-1}$) already constrain it. The predicted mass spectrum has the form of a broken power-law consisting of a pre-existing BH powerlaw mass spectrum and a harder powerlaw mass spectrum resulting from mergers. The predicted spin spectrum is multi-peaked with the evolution of retrograde spin BH in the gas disk playing a key role. We outline the large uncertainties in each of these LIGO observables for this channel and we discuss ways in which they can be constrained in the future.
\end{abstract}
\keywords{black holes, LIGO --- 
AGN disks --- mergers}
\section{Introduction} \label{sec:intro}
The gravitational wave (GW) events detected by the Advanced Laser Interferometer Gravitational-Wave Observatory (LIGO) correspond to the merger of stellar mass black holes (BH) considerably more massive than those observed in our own Galaxy. The upper end of the
range of BH merger rates derived from LIGO observations of 212~Gpc$^{-3}$~yr$^{-1}$ \citep{Abbott16b}
requires consideration of locations where BH mergers can occur faster than expected from 
GW emission alone. Among the first few LIGO detections are possible low value spin or misaligned spins, which may be problematic for models of binary evolution \citep{Oshaugh17}. While 
BHs with larger than expected masses 
can occur naturally in the field \citep{Belczynski10,deMink16},
they are more likely to form in regions with concentrations of BHs, such as galactic nuclear star clusters 
\citep{HopTal06,O'Leary09,AntRas16,Rodriquez16}. Massive gas disks in active galactic nuclei (AGN) provide natural locations for gas accretion and repeated mergers because the gas disk can drive migration of BH towards migration traps, reduce the inclination of intersecting orbits, enable binary formation, and harden existing binaries. Together, these effects can result in rapid increase in the mass of embedded BHs, potentially to observed values \citep[e.g.][]{McK12,McK14,Bello16,Bartos17,Stone17}.

In this paper we parameterize the expected merger rate, and the mass and spin distributions from this 
channel for comparison with the LIGO observations, and we discuss 
 how observations and simulations can constrain these predictions. 

\section{Model Outline}
\label{sec:drake}
Galactic nuclei likely contain some of the densest concentrations of BHs in the Universe \citep[e.g.][ and references therein]{Morris93,Miralda00}, so it is natural to look for BH mergers in galactic nuclei \citep{O'Leary09,McK12,Antonini14}. While BH binary mergers can occur at modestly enhanced rates (compared to the field) in nuclear star clusters just from dynamical binary hardening
\citep{AntRas16,Rodriquez16}, or capture from single-single \citep{O'Leary09} and binary-single encounters \citep{Samsing14}, a dense nuclear disk of gas can greatly accelerate the rate of BHB formation and merger \citep{McK12,McK14}.  

The simplest picture of this LIGO channel begins with a spherical distribution of BH, stars and other stellar remnants orbiting in the central $\rm{pc}^{3}$ of a galactic nuclei around a supermassive black hole (SMBH). Next, around the SMBH, we add a massive gas disk, which can be geometrically thin or thick. A fraction $f_{\rm co}$ of the initial number of BH in the nucleus $N_{BH}$, will have orbits coincident with the disk and approximately half of these orbits should be retrograde compared to the disk gas. Yet another fraction $f_{\rm g}$ of the population $N_{BH}$ intersect the disk on their orbits and are ground down into the plane of the disk within the AGN disk lifetime ($\tau_{\rm AGN}$). Thus an overall fraction $f_{d}=f_{\rm co}+f_{\rm g}$ of nuclear BH end up embedded in the disk, and quickly have their orbits damped and circularized by gas drag \citep[e.g.][]{McK12}. The net torques from disk gas causes BH to migrate within the disk and encounter each other at low relative velocities \citep{McK12,Bello16}. BH binaries that form in the disk are expected to merge efficiently due to gas torques \citep[e.g.][]{Haiman09,Kocsis11,McK11,Stahler10,Baruteau11,McK12}. BH mergers may preferentially occur in convergence zones containing migration traps \citet{Bello16} which occur in semi-realistic models of AGN disks \citep{Sirko03,Thompson05}. Multiple objects trapped in such orbits collide efficiently rather than being ejected (\citet{Horn12}; Secunda, Bellovary et al. (2018) in prep.). In this paper, we examine what constraints can be put on the merger rate and the BH spin and mass distributions for this AGN channel.

\section{Rate of black hole binary mergers in AGN disks}
\label{sec:rate}
We parameterize the rate of BH-BH mergers in AGN disks simply as:
\begin{equation}
{\cal R}=\frac{N_{GN} N_{BH} f_{AGN} f_{d}f_{b}\epsilon}{\tau_{AGN}}
\label{eq:rate}
\end{equation}
where $N_{GN} (\rm{Mpc}^{-3})$ is the average number density of galactic nuclei
in the Universe, $f_{AGN}$ is the fraction of galactic nuclei that have active AGNs which last for time $\tau_{AGN}$, $f_{d}=f_{co}+f_{g}$ is the fraction of 
nuclear BH that end up in the disk, $f_{b}$ is the fraction of BH
in BH-BH binaries in the disk, and $\epsilon$ represents the fractional change in number $N_{BH}$ of BH in the central 
region ($\sim \rm{pc}^{3}$) over a full AGN duty cycle \footnote{If $\epsilon \sim 1$ then $N_{BH}$ is
approximately conserved between AGN episodes. If $\epsilon (>)<1$ 
$N_{BH}$ (grows) shrinks between AGN phases due to the net effect of mergers, infall of new BH, stellar evolution etc..} ${\cal R}$ can be parameterized as:
  
\begin{eqnarray}
{\cal R}&=&12 \rm{Gpc}^{-3} \rm{yr}^{-1}\frac{N_{GN}}{0.006 \rm{Mpc}^{-3}} \frac{N_{BH}}{2 \times 10^{4}} \frac{f_{AGN}}{0.1} \\ \nonumber
&\times& \frac{f_{d}}{0.1}\frac{f_{b}}{0.1} \frac{\epsilon}{1}\left(\frac {\tau_{AGN}}{10\rm{Myr}}\right)^{-1}.
\label{eq:rate1}
\end{eqnarray}

However, if we want to constrain the constributions of this channel to LIGO observations, it is much more useful to show the allowed \emph{range} of ${\cal R}$ and the range of each of the contributing factors from eqn.~(\ref{eq:rate}), which we list in Table~\ref{tab:rates}. \\

The $N_{GN}$ lower limit corresponds to galaxies with stellar mass greater than or equal to that of the Milky Way \citep{Baldry12} as measured from Schechter function fits to galaxy luminosity functions \citep[e.g.][]{Cole01}. The $N_{GN}$ upper limit corresponds to dwarf galaxies with stellar mass $> 10^{9}\mbox{ M}_{\odot}$ \citep{Baldry12}, which includes all locally observed SMBH ($\geq 10^{5}\mbox{ M}_{\odot}$) inferred from $M-\sigma$ studies of galaxies and dwarf galaxies \citep{Reines15}.\\ 
Also in Table~1, $N_{BH} \sim 10^{3}$ corresponds to the number of BH allowed $\leq 0.1 \rm{pc}^{-3}$ of Sgr A* according to the distribution of the S-star orbits \citep{Antonini14}, whereas $N_{BH}\sim 10^{6}$~pc$^{-3}$ seems to be the maximal density allowed by simulations \citep{Antonini14}.\\
The lower limit to $f_{AGN}$ assumes only quasar disks are efficient BH merger sites and $f_{AGN} \sim 0.3$ assumes all LINER galactic nuclei \citep{Ho08} consist of advection dominated accretion flows (ADAFs) with high accretion rate \citep{PW80,Narayan95,Lasota16}, capable of driving BH mergers. \\
The binary fraction of BH $f_{b}$ has been estimated to be as high as $f_{b} \sim 0.2$ \citep{Antonini14}), but dynamically hot environments such as star clusters, could actually yield very low binary fractions $f_{b} \leq 0.01$ over time in the absence of gas \citep{miller12,leigh16} due to the large number of 'ionizing' interactions, so we choose $f_{b}=[0.01,0.2]$ in Table~1.\\ 
Reasonable estimates of $\tau_{AGN}$ span 0.1-100Myr \citep{Haeh93,King15,Schawinski15}. $\cal{R}$ will be highest if AGN episodes are short-lived but frequently repeated and efficient at BH mergerse. These circumstances ensure that there are multiple opportunities for BH in a galactic nucleus to encounter each other at low relative velocity and merge in a disk.

\begin{deluxetable}{c|ccc}
\tablecaption{Parameter ranges in Eqn.~\ref{eq:rate}. \label{tab:rates}}
\tablehead{
\colhead{Parameter} & \colhead{Lower} & \colhead{Upper} &
}
\startdata
$N_{GN}^{a}$(Mpc$^{-3}$)  &$4\times10^{-3}$&$10^{-2}$ &\\
$N_{BH}^{b}$(pc$^{-3}$) &$10^{3}$ & $10^{6}$ & \\
$f_{AGN}^{c}$ &0.01&0.3 &\\
$f_{b}$ &0.01&0.2 &\\  
$f_{d}^{d}$ &0.01&0.7 &\\
$\tau_{AGN}$(Myr) &1 &100 &\\
$\epsilon$ &0.5&2 &\\
${\cal R}$(Gpc$^{-3}${~}yr$^{-1}$)  &$10^{-4}$  &$10^{3}$ &\\
\enddata
\tablecomments{Range of parameters in Eqn.~(\ref{eq:rate}) and range of merger rate (see text). $^{a}$ from \citet{Baldry12}. $^{b}$ from \citet{Miralda00,Antonini14}. $^{c} f_{AGN} \sim 0.1$ for Seyfert AGN \citep{Ho08}. $f_{AGN} \sim 0.3$ with all
LINERs and other low luminosity AGNs. $^{d}$ $f_{d}=f_{co}+f_{g}$. $f_{co}$ comes from h/R, the disk aspect ratio. h/R $\sim$0.01--0.1 \citep{Sirko03}. $h/R \sim 10^{-3}$--
$0.1$ \citep{Thompson05}. $h/R \sim 0.1$--0.7 in super-Eddington ADAFs \citep{Lasota16}. $f_{g}$ depends on $h/R$, $\rho_{\rm disk}$ and $\tau_{AGN}$.}
\end{deluxetable}

From Table~1, the allowed range from Eqn.~(\ref{eq:rate}) is ${\cal R} \sim 10^{-4}$--$10^{4}\mbox{ Gpc}^{-3} \mbox{ yr}^{-1}$.  The upper bound to the LIGO BH binary merger rate of $\sim 240 \mbox{Gpc}^{-3} \mbox{ yr}^{-1}$ already rules out upper limits to most
parameters in Table~1 \footnote{The LIGO rate upper bound places a \emph{lower limit} on $\epsilon$, since a small value of $\epsilon$ suggests most BH in AGN are consumed in mergers and would imply a much greater $\cal{R}$ than observed} and \emph{allows actual astrophysical limits to be placed on models of AGN disks by LIGO BH merger detections}. Future observational
constraints and simulation results will, however, be required to figure out
\emph{which} upper limits are ruled out by LIGO. For example, the upper limit to $N_{GN}$ could be reduced by contrasting activity rates as a function of galactic mass in a complete sample. The inferred $N_{BH}$ can be constrained via population studies of the X-ray emission from binaries around Sgr A* and in M31, as well as via dynamics studies of the number density of BH allowed from the orbital parameters of stars in galactic nuclei. The upper limit on $f_{AGN}$ can be reduced if we can observationally distinguish between high- and low-accretion rate LINERs.   Simulations that include a spherical component of individual stars and BH as well as migrating objects in the disk are required to properly constrain $f_{b}$. Encounters between objects from the spherical dynamical
component and the disk dynamical component will occur at relatively
high velocity and can therefore ionize sufficiently soft, large radius, binaries. Thus, in order for $f_{b}$ to be moderately large in this channel, we require $f_{g}$ to be large, since otherwise the rate of ionizing encounters can ionize binaries \citep{Leigh17}. So limits on $f_{g}$ from semi-analytic approaches or simulations \citep{Kennedy16} can also help constrain $f_{b}$.

Uncertainties in $\cal{R}$ are
dominated mainly by lack of knowledge of the distribution and number of BH in galactic nuclei, how efficiently gas disks can grind down orbits, and whether geometrically thick disks can
efficiently merge BHs. Understanding multiple-object migration 
and the role of retrograde orbiters is another key area for future work.

\section{Constraining BH masses}
\label{sec:masses}
By merging BHs in AGN disks, we expect 'overweight' BH to result \citep{McK12}. To investigate the range of BH masses involved in mergers in this channel, we use a toy model calculation of the evolution of a population of  BH embedded and migrating in an AGN disk. We made many simplifying assumptions: there are no BH binaries to begin with ($f_{b}=0$), BH remain in the disk after merger, tertiary encounters are neglected, no BHs merge with the SMBH, no new BH are added to the population ($f_{g}=0$) and we ignore mass growth due to gas accretion. We began with a uniform distribution of BH drawn from a \citet{kroupa} initial mass function 
$N_{BH}(M) \propto M^{-\gamma_{0}}$, with $\gamma_{0}=2.3$ distributed over three mass bins (5,10,15$M_{\odot}$) and chose normalization $N_{BH}(5\mbox{ M}_{\odot})=10^{3}$. 

A BH on a prograde orbit in an AGN disk with mass $M_1$ will migrate on a (Type~I) timescale \citep{Paarde10, McK12} 
\begin{eqnarray}
t_{\rm mig} &\approx& 38\rm{Myr} \left(\frac{N}{3}\right)^{-1}\left(\frac{R_{b}}{10^{4}r_{g}}\right)^{-1/2}\left(\frac{M_{1}}{5\rm{M}_{\odot}}\right)^{-1} \nonumber \\
&\times&\left(\frac{h/R_{b}}{0.02}\right)^{2}\left(\frac{\Sigma}{10^{5}\rm{kg m}^{-2}}\right)^{-1}\left( \frac{M_{\rm SMBH}}{10^{8} {\rm{M}_{\odot}}} \right)^{3/2}
\label{eq:mig}
\end{eqnarray}
where $N$ is a numerical factor of order $3$. So the toy model population outlined above will evolve over time. If $10^{3}$ BH are uniformly distributed across a disk of radius $R_d\sim 10^{5}r_{g} ,(r_{g}=GM_{\rm SMBH}/c^{2})$, BH orbits are separated by $\sim 10^{2}r_{g}$ on average. This separation could be closed in $\sim 0.4$~Myr from eqn.~(\ref{eq:mig}). Our initial distribution of singleton BH separated by $\sim 10^{2}r_{g}$ on average will therefore evolve from $f_{b}=0$ towards $f_{b} \sim 0.5$ within $\sim 0.4$Myr due to migration. The probability of encounter between BH of masses $M_{1}, M_{2}$ in time $\Delta t$ is
\begin{equation} \label{eq:prob}
P(M_{1}|M_{2}) \propto \frac{N(M_{1})N(M_{2})}{t_{\rm mig}(M_{1})
t_{\rm mig}(M_{2})}.
\end{equation} 

When a pair of BHs approaches within their binary Hill radius $R_H =(q/3)^{1/3}R_{b}$, where $q$ is the binary mass ratio and $R_{b}$ is the radius of the binary center of mass, gas drag can cause them to merge rapidly. \citet{Baruteau11} showed that binary semi-major axis $a_{b}$ halves due to gas drag in only 200(1000) orbits about the binary center of mass for a retrograde (prograde)
binary compared to gas velocity. Using this result, a BH binary with $a_{b}=R_{H}$ at $R_{b} \sim 10^{3}r_{g}$ has a
characteristic timescale for binary hardening of 0.4~kyr (8~kyr) in
the retrograde(prograde) case. Only 20--25 such halvings (corresponding to $\sim 0.1$--0.2~Myr, naively assuming a constant gas hardening rate) would shrink $a_{b}$ sufficiently that GW emission takes over and the merger happens promptly. The gas hardening rate may be even faster than this estimate since more gas enters the binary's Hill sphere as it shrinks \citep{Baruteau11}, which may pump binary eccentricity. However, gas torques may decrease in efficiency once the binary has hardened sufficiently that the binary velocity is substantially supersonic compared to most gas within the Hill radius \citep{Sanchez14}. For our toy model, we therefore assume $\sim 0.1$Myr is the minimum gas hardening timescale to merger, but we note that the actual gas hardening timescale could take up to an order of magnitude longer. 

In our toy model, if the typical time for a BH to encounter another BH in the disk is $\sim 0.4$Myr, then adding an additional $\sim 0.1-1$Myr for a gas-hardening timescale, yields a characteristic time to merger of $\sim 0.5-1.5$Myr in our model. So, we expect that around half the initial population of our toy model will have encountered each other and merged in this time. In calculating the evolution of our toy model, we chose $\Delta t \sim 0.1-0.3$Myr to correspond to a time when $\sim 10\%$ of the initial population of lowest mass BHs ($5\mbox{ M}_{\odot}$) have encountered each other and merged. All other encounters are normalized to this encounter rate. For simplicity, we assume all binaries formed in $\Delta t$ merge within that time, and we neglect the mass-energy loss from the mergers. After $\Delta t$, all BH that merged are removed from their original mass bins, and the newly merged object is added to the appropriate mass bin. 

\begin{figure}
\centering
\begin{center}
\includegraphics[width=6.0cm,angle=-90]{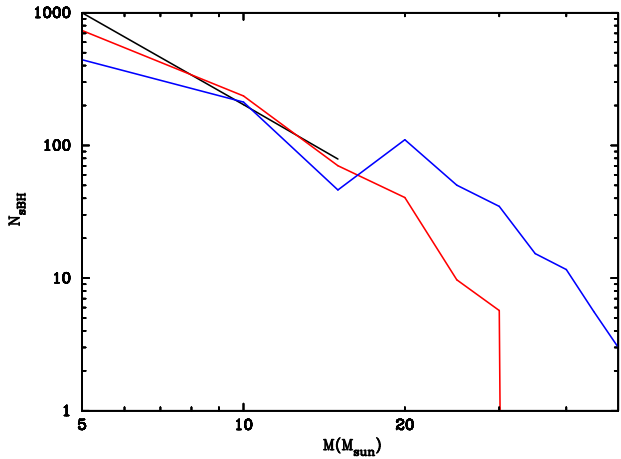}
\end{center}
\caption{
Evolution of an initial $5-15M_{\odot}$ BH mass distribution (black curve)
in an AGN disk based on a toy merger model. Black curve corresponds to the initial BH mass distribution. Red and
blue curves shows the evolution of the distribution after timesteps
corresponding to $\Delta t \approx 0.1-0.3$~Myr and $\Delta t^{\prime} \sim 0.2-0.6$~Myr respectively (see text). A choice of heavier inital mass range will alter upper mass limits.  
\label{fig:nsbh}}
\end{figure}

Figure~\ref{fig:nsbh} demonstrates the simplistic evolution expected as the initial BH distribution (black line) evolves to the red curve in time step $\Delta t \sim 0.1-0.3$Myr, where $\sim 10\%$ of the lowest mass BHs in the initial (black) distribution have merged. The red curve evolves to the blue curve after an additional $\Delta t^{\prime} \sim 0.2-0.6$Myr, when $\sim 10\%$ of the lowest mass BH on the red curve are expected to merge. The BH mass distribution in our toy model flattens from $\gamma_0 = 2.3$ to $\gamma \sim 2$ as low-mass BH are consumed.

Now assume that BH from the non-disk spherical population, interact with the disk and their orbits are ground down into the disk, i.e. $f_{g}>0$. The addition of some of the (initially) spherical BH population into the disk will support the BH mass distribution in the disk at the low mass end. So an initial power law distribution $\propto M^{-\gamma_{0}}$ of BH mass will evolve towards a broken-power law distribution of the form
\begin{equation} \label{eq:dist}
N_{BH} \propto \left\{ 
     \begin{array}{l} 
       N_{1}M^{-\gamma_{1}} \mbox{ for }M<M_{\rm break} \\ 
       N_{2}M^{-\gamma_{2}} \mbox{ for } M>M_{\rm break} 
     \end{array}, 
   \right.
\end{equation}
where $\gamma_{2} < \gamma_{1}$, $N_{1}/N_{2} \sim  (f_{g}/f_{co})$, where $f_{co}$ is the fraction of BH initially in the disk and on average $f_{g}$ is the fraction of BH ground down into the disk over $\tau_{AGN}/2$ and $M_{\rm break}$ lies near the upper end of the inital mass range ($\sim 15
\mbox{ M}_{\odot}$ in our toy model). 

In order to include gas accretion in this toy model, we assumed a gas accretion rate for BH on [retrograde, prograde] orbits of
$\dot{M}_{1} \sim [10^{-2},1] \dot{M}_{\rm Edd}$, where
 \begin{eqnarray}
\dot{M}_{\rm Edd}&=&\frac{4\pi G M_{1} m_{p}}{\eta c} \nonumber \\
 &\approx& 2.2 \times 10^{-7} \frac{M_{\odot}}{\rm{yr}} \left(\frac{\eta}{0.1}\right)^{-1}\left( \frac{M_{1}}{10M_{\odot}}\right)
\label{eq:medd}
\end{eqnarray}
 is the Eddington mass accretion rate with $m_{p}$ the proton mass and $\eta$ the accretion luminosity  efficiency. Over an AGN disk lifetime of $\tau_{AGN} \sim 10$Myr, we can neglect gas accretion onto BH on retrograde orbits. 

\begin{deluxetable}{c|cc}
\tablecaption{Parameter ranges in BH masses. \label{tab:mass}}
\tablehead{
\colhead{Parameter} & \colhead{Lower} & \colhead{Upper} 
}
\colnumbers
\startdata
$M_{b} (\mbox{ M}_{\odot})$($\gamma=2$) & 10& 100 \\
$M_{b} (\mbox{ M}_{\odot})$($\gamma=1$) & 10& 500 \\
$M_{b} (\mbox{ M}_{\odot})$($\gamma=$broken) & 10& 500 \\
$q$($\gamma=2$)          &0.1 & 1   \\
$q$($\gamma=1$)          &0.01 & 1  \\
$q$($\gamma=$broken)          &0.01 & 1  \\
\enddata
\tablecomments{Parameter ranges predicted for BH binaries in this channel, assuming initial BH mass range 5--15~M$_{\odot}$ and uniform distribution of BH (see text). }
\end{deluxetable}

In Table~\ref{tab:mass} we list parameter ranges for BH masses on the basis of the probabilistic toy model outlined above for three different assumptions: 1) $N_{BH} \propto M^{-2}$ (roughly the blue curve in Fig.~\ref{fig:nsbh}), corresponding to a short lived disk with $f_{\rm co} \gg f_{\rm g}$. 2) $N_{BH} \propto M^{-1}$, corresponding either to a long lived disk ($\tau_{AGN}>10$Myr) or efficient gas hardening with a low rate of orbit grind down ($f_{\rm co} \gg f_{\rm g}$). 3) $N_{BH} \propto M^{-2}(M^{-1.5})$ for $M<15M_{\odot}(>15M_{\odot})$, corresponding either to efficient orbit grind down ($f_{\rm g} \sim f_{\rm co}$), or efficient stellar formation and evolution in the disk with a new top-heavy IMF. In Table~\ref{tab:mass} we list the binary mass ratio $M_{b}$ range for each set of assumptions. The lower limit to $M_{b}$ is trivially the lowest possible mass binary drawn from the initial mass distribution, with no growth from gas accretion and the upper limit to $M_{b}$ is simply the highest mass binary in the distribution. Also listed in Table~\ref{tab:mass} are the range of mass ratios ($q$) of the binaries in the three different scenarios, with the lower limit given by the range of BH masses allowed in the three different distributions and $q=1$ is the trivial upper limit.\\

If the fraction of BH ground down
into the disk $f_{\rm g}(t) \geq f_{\rm co}(t)$, the fraction of BH
coincident with the disk, which will be true for relatively
long-lived, thin ($h/R \ll 1$) disks, the BH mass spectrum evolves from an initial power-law distribution to a broken power-law as in Eqn.~(\ref{eq:dist}) with $\gamma_{1} \sim \gamma_{0} > \gamma_{2}$. The uncertainty in mass estimates for this channel is driven mainly by the initial mass distribution of BH in
the central region, as well as the ratio of $f_{\rm g}(t)/f_{\rm co}(t)$, which in turn depends on disk density and $h/R$. 

\section{Range of  BH spins}
\label{sec:spins}
As black holes in the AGN disk accrete gas and merge with each other, their initial spin distribution will change with time.
Assuming a uniform distribution of spins ($a$) and angular momenta ($L$) for BH in galactic nuclei, there will be \emph{four} distinct populations of BHs in AGN disks as follows: 
\begin{enumerate}
\item Prograde spin, on prograde orbits, denoted by ($a^{+},L^{+}$). 
\item Prograde spin, on retrograde orbits ($a^{+},L^{-}$).
\item Retrograde spin, on prograde orbits ($a^{-},L^{+}$).
\item Retrograde spin, on retrograde orbits ($a^{-},L^{-}$).
\end{enumerate}
We expect the fraction $f_{\rm co}$ of BH co-orbital with the AGN disk should have an initial uniform distribution across all four BH populations.

The four BH populations will evolve differently due to gas accretion. The ($a^{+},L^{+}$) population rapidly accretes gas, spins up, and aligns spins with the disk gas once the BH has accreted a few $\%$ of its own mass\citep{Bogdanovic07}, i.e. in $< \tau_{AGN}$. An initially uniform spin distribution $a^{+}=[0,+0.98]$ evolves towards $a^{+} \sim 0.98$ at an average rate $\sim (\tau_{AGN}/40\rm{Myr})(\dot{m}/\dot{M}_{\rm Edd})$ where $\dot{m}/\dot{M}_{\rm Edd}$ is the average gas accretion rate as a fraction of the Eddington rate (which takes $\approx 40$Myr to double mass). By contrast, the ($a^{+},L^{-}$) population faces a strong headwind, so it accretes very weakly from the gas. An initially uniform distribution of spins in this population will remain uniform over $\tau_{AGN}$. The ($a^{-},L^{+}$) population spins down towards $a \sim 0$ after an increase of mass by a factor $\sqrt{3/2}$ \citep{Bardeen70} and will then join the ($a^{+},L^{+}$) population. The ($a^{-},L^{-}$) population spins down more slowly due to the headwind and so an initial uniform distribution of spins remains uniform over $\tau_{AGN}$.

BH mergers will further complicate the spin evolution of the four BH populations. The four populations interact due to migration and form binaries if captured within the binary Hill sphere. Binary orbital angular momentum ($L_{b}$) is the dominant contributor to the spin of the merged BH binary so equal mass BH mergers yield merger products with $|a|\sim 0.7$ \citep{Hofmann16}. Binaries can form with prograde or retrograde
orbital angular momentum compared to the disk gas (denoted by
$L_{b}^{\pm}$). If a binary forms with retrograde orbital angular momentum
($L_{b}^{-}$),  the merger is faster than in the prograde case
\citep{Baruteau11}, and the merger product will have $a^{-}=-0.7$ (i.e. retrograde spin
compared to disk gas). Thus the fastest growing of the four populations of BH in the disk due to mergers will actually be $(a^{-},L^{\pm})$. This population evolves towards low spin ($a \sim 0$) due to gas accretion, at an average rate $\sim (\tau_{AGN}/40\rm{Myr})(\dot{m}/\dot{M}_{\rm Edd})$. Among the initial fraction $f_{co}$ of co-orbital BHs, we expect equal numbers of prograde to retrograde orbits. However, since prograde orbits are ground down faster (smaller headwind, greater Bondi radius), we expect $(a^{\pm},L^{+})/(a^{\pm},L^{-}) \approx 1+(f_{g}/f_{co})$.

\begin{figure}
\centering
\begin{center}
\includegraphics[width=6.0cm,angle=-90]{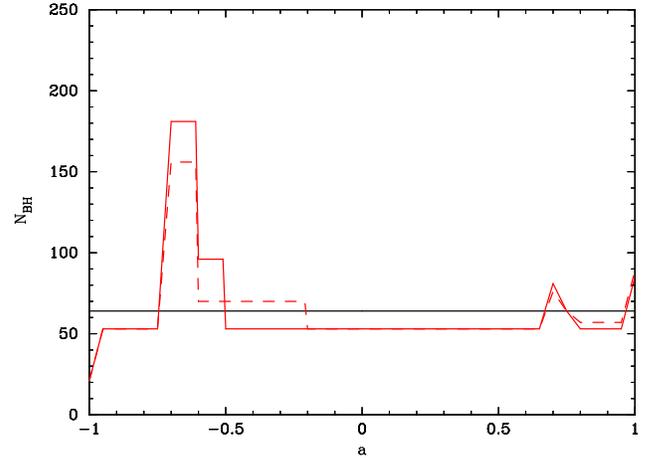}
\end{center}
\caption{
Evolution of an initial uniform BH spin distribution 
in an AGN disk based on a toy merger model, including gas accretion (see text). Spins are binned per 0.05 of spin parameter ($a$). Black line corresponds to a uniform BH spin distribution for the initial population. The corresponding initial mass distribution is given by the black curve in Fig.~\ref{fig:nsbh}. The red solid curve shows the spin distribution after the toy model has evolved for $\Delta t \approx 0.1-0.3$~Myr to include mergers and gas accretion at the Eddington rate. The corresponding mass distribution after this time is given by the red curve in Fig.~\ref{fig:nsbh}. The red dashed curve is as the solid curve, except we assume super-Eddington accretion at $\times 5$ the Eddington rate.
\label{fig:spin}}
\end{figure}

Applying all of this to our toy model above allows us to construct the spin distribution in Fig.~\ref{fig:spin}. An initial uniform spin distribution (black line) evolves towards the solid red curve after $\Delta t \approx 0.1-0.3$Myr. The corresponding mass distribution is the red curve in Fig.~\ref{fig:nsbh}. The red solid curve in Fig.~\ref{fig:spin} shows a prominent peak at $a=-0.7$ due to a $\times 5$ faster merger rate of retrograde binaries and a smaller peak at $a=+0.7$ due to mergers of prograde binaries. Both peaks are smeared out towards the right by gas accretion during $\Delta t$ and will consist of BH masses $\geq 10M_{\odot}$ from the initial mass distribution. Some pile-up is happening at $a>0.95$ due to gas accretion onto the already near maximal spinners of the ($a^{+},L^{+}$) population. The red dashed curve shows what happens if we assume gas accretion can occur at super-Eddington rates onto BH in the disk ( $\times 5$ the Eddington rate). In particular the 
more massive merged population at $a \sim -0.7$ gets quickly smeared out and driven towards low spin. Thus, from Fig.~\ref{fig:spin} if LIGO  constrain the spins of most merger precursor BHs to be small, the AGN channel requires super-Eddington accretion onto initially retrograde spin BH to grow this population.

Only the ($a^{+},L^{+}$) population will align or anti-align relatively quickly with the AGN disk gas. Assuming the ($a^{+},L^{+}$) population are all aligned or anti-aligned with the disk gas, by drawing randomly from a uniform distribution across ($a^{\pm},L^{\pm}$), there is a $\approx 1/16$ chance that both BH have (anti-)aligned spins and represents our lower limit for the fraction of BH (anti-) aligned with disk gas. If $f_{\rm g}(t) \gg f_{\rm co}(t)$, then effectively the two populations ($a^{\pm},L^{+}$) will dominate so  $f_{\pm \rm align} \approx 1/4$, which is our approximate upper limit for the fraction of BH (anti-) aligned with disk gas. Our estimates of $f_{\pm \rm align}$ suggest that a larger population of mergers will be requied to test this channel in population spin studies than estimated by \citet{Fishbach17,Gerosa17}. Anti-aligned binaries in the AGN disk allow LIGO a unique chance to test the spin precession instability \citep{Gerosa15}.

Once a BH binary merges, the resulting merger product can experience a
gravitational radiation recoil kick of $v_{\rm kick} \sim20$--400~km~s$^{-1}$, depending on relative spins and mass ratios
\citep[e.g.][]{Merritt04,Campanelli07}. The result of  kicks from
  mergers between aligned and anti-aligned objects is to incline the merger product's
orbit relative to the AGN disk by 
$\theta=\tan^{-1}(\rm{v}_{kick}/\rm{v}_{\rm orb})$ where $v_{\rm orb}$ is the orbital velocity of the binary center of mass. Since $v_{\rm orb} \gg 400$km/s in most of the disk, the orbital inclination perturbation is at most a few degrees and the merger
product could be ground back down into the disk in time $<\tau_{AGN}$. Mergers of BH with spins out of alignment with the plane
of the disk and each other can produce the largest magnitude kicks (up
to several thousand kilometers per second) \citep[e.g.][]{SchnittBuon07,Lousto12}. Such mergers will be rare, but will produce large kicks ($\propto q^{2}/(1+q)^{4}$ in the mass ratio $q$, \citet{Campanelli07}), escape the disk at angle $\theta$ and may not be ground back down within $\tau_{AGN}$.

Table~\ref{tab:spin} summarizes the ranges allowed for spins in this LIGO channel. The typical spin distribution depends on the relative fractions of the four populations of BH in the disk ($a^{\pm},L^{\pm}$) and their evolution as $f_{g}/f_{co}$ changes, driven in turn by disk aspect ratio ($h/R$) and the disk gas density and $\tau_{AGN}$.  We expect an initial population uniform across  ($a^{\pm},L^{\pm}$), but ($a^{\pm},L^{+}$) will grow with the fraction $f_{\rm g}(t)$ of BH ground-down into the disk. Peaks will arise in the spin distribution at $a\sim -0.7,+0.7$ due to mergers and gas accretion will drive $a^{-} \rightarrow 0$ and $a^{+} \rightarrow 0.98$ independent of mergers. Gas accretion at super-Eddington rates plus faster mergers by retrograde binaries may be required to generate a population of overweight, low spin BH in the AGN disk. 

\begin{deluxetable}{c|cc}
\tablecaption{Parameter ranges in BH spins. \label{tab:spin}}
\tablehead{
\colhead{Parameter} & \colhead{Lower} & \colhead{Upper} 
}
\colnumbers
\startdata
$a^{+} (L^{+})$    &0 &0.98 \\
$a^{-} (L^{+})$    &-0.98 &0  \\
$a^{+} (L^{-})$    &0.0 &0.98  \\
$a^{-} (L^{-})$    &-0.98 &0  \\
$a_{\rm merge}$    &-0.7 &+0.7  \\
$f_{\pm \rm align}$ & 0.06&0.25  \\
\enddata
\tablecomments{Parameter ranges allowed for BH spins in this channel (see text). }
\end{deluxetable}

\section{Observational Constraints: GW}
\label{sec:imp_gw}
Binary black hole mergers in an AGN disk imply unique, testable
predictions that would not be expected from other BH
merger channels, including: 1. A spin distribution (see \S\ref{sec:spins}) that includes aligned/anti-aligned spin binaries and 2. a population of overweight BH or IMBH orbiting SMBHs, generating GWs
detectable with the Laser Interferometer Space Antenna (LISA) \citep{McK14}. 
A circularized IMBH-SMBH binary at a migration trap ($a_{b} \sim 10^{2}r_{g}$) around a SMBH with
$M_{\rm SMBH}<10^{7}\mbox{ M}_{\odot}$ will be detectable with LISA at
modest signal-to-noise ratio in a year's observation
\citep{McK14}. If AGN disks are efficient at gas-driven mergers of BH, we expect that every AGN must contain one or more IMBH-SMBH binaries, implying an approximate rate comparable to that in \citet{PortegiesZ06}. 

\section{Observational Constraints:EM}
The brightest AGN are too bright compared to any short-term 
EM signal that might result from a BH merger in a
gas disk. Low luminosity AGN might permit short timescale EM events from BH mergers to be visible.  As IMBHs grow in
migration traps, gaps and cavities in the accretion flow can form 
and oscillations on the dynamical timescale of the accreting IMBH can
be detected in optical, UV, and X-ray spectral signatures
\citep[e.\ g.][]{McK13,McK14,McK15}. Temporal and energetic
asymmetries in the X-ray signatures are best detected using
micro-calorimeters, such as the one that will fly on the X-ray Astronomy Recovery Mission succeeding
 \emph{Hitomi}. Perturbations of the innermost disk will occur as migrators in the disk plunge into the SMBH and temporarily dominate the local co-rotating mass, detectable in large UV-optical quasar surveys \citep{Drake09} as well as the X-ray band. Large optical surveys of quasar disks can also limit total supernova rates due to migrating/accreting/colliding stars \citep{Graham17}, in turn placing limits on the disk populations of stars and stellar remnants. Estimates of the rates of transits by bloated stars, best detected in the X-ray band \citep{McK98}, can put limits on the population on spherical orbits around and passing through AGN disks.

As the AGN phase ends, remaining BH will interact dynamically, so the distribution of orbital parameters of the
BHs and stars entrained in the disk will relax. \citet{Alex07} show that
if very massive stars ($>10^{2}M_{\odot}$) exist in our own Galactic nucleus, they can pump the
eccentricity distribution of massive stars to even $e \sim 0.4$ within 5~Myrs. However, such stars are
short-lived and observed stellar eccentricities reach $e \sim 0.7$
\citep{Paumard06}. On the other hand, a population of overweight BHs caused by merger in
an AGN disk can rapidly pump stellar orbital eccentricites post-AGN and inflate the thickness
($h/R$) of stellar disks in galactic nuclei. Thus, if this BH merger
channel is efficient, thin disks of stars will not be observed in
post-AGN galactic nuclei.   

Neutron stars (NS) should also exist in AGN disks, and can migrate. So there should be a correlation between NS-NS and NS-BH mergers in AGN disks and the rate of BH-BH mergers expected from this channel.
No correlation has been observed so far between short gamma-ray bursts in the
local universe and AGNs \citep{Berger14}, but so far, only a handful of short
gamma-ray bursts have sufficiently accurate positions in the sky to rule
out an association with AGN in these cases. The efficiency of this
LIGO channel could be further constrained by ongoing 
studies of the correlation of short gamma-ray bursts with AGN. Future simulations could usefully 
focus on the expected distribution of NS in mass segregating clusters
in galactic nuclei, and ultimately on determining the expected NS
merger rate in AGN disks.

\section{Conclusions}
\label{sec:conclusions}
We parameterize the rate of black hole mergers within AGN disks and the mass and spin distributions that result. The strongest observational constraints can be placed on this channel by: 1. ruling out a population of maximal spin BH via LIGO, 2. ruling out a correlation betwen short gamma-ray bursts and AGN, 3. constraining the rate of obscured supernovae in AGN
disks via studies of large samples of AGN, 4. ruling out a population of high
accretion rate ADAFs in galactic nuclei and 5. observing very thin disks of stars in nearby Galactic nuclei. Future simulations should focus on 1. the ratio of NS/BH in nuclear star clusters
undergoing mass segregation, 2. encounters between prograde and
retrograde orbiters in AGN disks and 3. interactions and binary
formation between BHs with pro- and retro-grade spins and orbits at migration traps in a range of AGN disk models.  If AGN are efficient at merging BH, LISA will detect a large population of IMBH in disks around SMBH in the nearby Universe. 

{\section{Acknowledgements.}} Thanks to Maya
Fishbach, Davide Gerosa, Matthew Graham, Daniel Holz, Dan Stern and Nick Stone for useful
conversations. BM \& KESF are supported by NSF PAARE AST-1153335 and NSF PHY11-25915. BM \& KESF thank CalTech/JPL and NASA GSFC for support during sabbatical. M-MML is partly supported by NSF AST11-09395.




\begin{thebibliography}{}
\bibitem[Abbott et al. (2016a)]{Abbott16a} Abbott B.P. et al., 2016, PhRvL, 116,1102
\bibitem[Abbott et al. (2016b)]{Abbott16b} Abbott B.P. et al., 2016, ApJL, 833,L1
\bibitem[Antonini et al. (2014)]{Antonini14} Antonini F., 2014, ApJ, 794, 106
\bibitem[Antonini \& Rasio (2016)]{AntRas16} Antonini F. \& Rasio F., 2016, ApJ (submitted), arXiv:1606.04889
\bibitem[Alexander et al. (2007)]{Alex07} Alexander R.D., Begelman M.C. \& Armitage P.J., 2007, ApJ, 654, 907
\bibitem[Baldry et al. (2012)]{Baldry12} Baldry I.K. et al., 2012, MNRAS, 421, 621 
\bibitem[Bardeen (1970)]{Bardeen70} Bardeen J.M., 1970, Nature, 226,64 
\bibitem[Bartos et al. (2017)]{Bartos17} Bartos I., Kocsis B., Haiman Z. \& M\'{a}rka S., 2017, ApJ, 835,165 
\bibitem[Baruteau et al. (2011)]{Baruteau11} Baruteau C., Cuadra J. \& Lin D.N.C., 2011, ApJ, 726, 28
\bibitem[Bellovary et al. (2016)]{Bello16} Bellovary J., Mac Low M.-M., McKernan B. \& Ford K.E.S., 2016, ApJ, 819, L17
\bibitem[Belczynski et al. (2010)]{Belczynski10} Belczynski K., Bulik T., Fryer C.L., Ruiter A., Valsecchi F., Vink J.S. \&  Hurley J.R., 2010, ApJ, 714, 1217
\bibitem[Berger (2014)]{Berger14} Berger E., 2014, ARA\&A, 52, 43
\bibitem[Bogdanovic et al. (2007)]{Bogdanovic07} Bogdanovic T., Reynolds C.S. \& Miller M.C., 2007, ApJ, 661, L147
\bibitem[Campanelli et al. (2007)]{Campanelli07} Campanelli M., Lousto C.O., Zlochower Y. \& Merritt D., 2007, PhRvL, 98, 231102
\bibitem[Cole et al. (2001)]{Cole01} Cole S. et al., 2001, MNRAS, 326, 255
\bibitem[Drake et al. (2009)]{Drake09} Drake A.J. et al., 2009, ApJ, 696, 870
\bibitem[deMink \& Mandel (2016)]{deMink16} deMink S. \& Mandel I., 2016, MNRAS, 460, 3545
\bibitem[Fishbach et al. (2017)]{Fishbach17} Fishbach M., Holz D.E. \& Farr B., 2017, ApJ, 840, L24
\bibitem[Gerosa et al. (2015)]{Gerosa15} Gerosa D. et al., 2015, Phys Rev Lett, 115, 141102
\bibitem[Gerosa \& Berti (2017)]{Gerosa17} Gerosa D. \& Berti E., 2017, Phys Rev D (submitted), arXiv:1703.06223 
\bibitem[Graham et al. (2017)]{Graham17} Graham M. et al., 2017, MNRAS, submitted
\bibitem[Haehnelt \& Rees (1993)]{Haeh93} Haehnelt M.G. \& Rees M., 1993, MNRAS, 263, 168
\bibitem[Haiman et al. (2009)]{Haiman09} Haiman Z., Kocsis B. \& Menou K., 2009, ApJ, 700, 1952
\bibitem[Ho (2008)]{Ho08} Ho L.C., 2008, ARA\&A, 46, 475 
\bibitem[Hofmann et al. (2016)]{Hofmann16} Hofmann F., Barausse E. \& Rezzolla L., 2016, arXiv:1605.01938 
\bibitem[Hopman \& Alexander (2006)]{HopTal06} Hopman C. \& Alexander T., 2006, ApJ, 645, L133 
\bibitem[Horn et al. (2012)]{Horn12} Horn B., Lyra W., Mac Low M.-M. \& S\'{a}ndor Z., 2012, ApJ, 750, 34 
\bibitem[Kennedy et al. (2016)]{Kennedy16} Kennedy G. et al., 2016, MNRAS, 460, 240
\bibitem[King \& Nixon (2015)]{King15} King A. \& Nixon C.J., 2015, MNRAS, 453, L46
\bibitem[Kocsis et al. (2011)]{Kocsis11} Kocsis B., Yunes N. \& Loeb A., 2011, PRD, 84, 024032
\bibitem[Kroupa (2002)]{kroupa} Kroupa P. 2002, Science, 295, 82
\bibitem[Lasota et al. (2016)]{Lasota16} Lasota J.-P. et al., 2016, A\&A, 587, 13  
\bibitem[Leigh et al. (2016)]{leigh16} Leigh
  N. W. C., Antonini F., Stone N. C., Shara M. M., Merritt D. 2016,
  MNRAS, 463, 1605
\bibitem[Leigh et al. (2017)]{Leigh17} Leigh
  N. W. C., Geller, A. M., McKernan, B., Ford, K. E. S., Mac Low,
  M.-M., Bellovary, J., Haiman, Z., Lyra, W., Samsing, J., O'Dowd, M.,
  Kocsis, B., Endlich, S. MNRAS, submitted (ArXiv:TBD)
\bibitem[Lousto et al. (2012)]{Lousto12} Lousto C.O., Zlochower Y., Dotti M. \& Volonteri M., 2012, PRD, 85, 084015
\bibitem[McKernan \& Yaqoob (1998)]{McK98} McKernan B. \& Yaqoob T.,  1998, ApJ, 501, L29
\bibitem[McKernan et al. (2011)]{McK11} McKernan B. et al, 2011, MNRAS, 417, L103
\bibitem[McKernan et al. (2012)]{McK12} McKernan B., Ford K.E.S., Lyra W. \& Perets H.B., 2012, MNRAS, 425, 460
\bibitem[McKernan et al. (2013)]{McK13} McKernan B., Ford K.E.S., Kocsis B. \& Haiman Z., 2013, MNRAS, 432, 1468
\bibitem[McKernan et al. (2014)]{McK14} McKernan B., Ford K.E.S., Kocsis B., Lyra W. \& Winter L.M., 2014, MNRAS, 441, 900
\bibitem[McKernan \& Ford (2015)]{McK15} McKernan B. \& Ford K.E.S., 2015, MNRAS, 452, L1
\bibitem[Miller \& Davies (2012)]{miller12} Miller M. C., Davies M. B. 2012, ApJ, 755, 81 
\bibitem[Merritt et al. (2004)]{Merritt04} Merritt D., Milosavljevi\'{c} M., Favata M., Hughes S.A. \& Holz D.E., 2004, ApJ, 607, L9
\bibitem[Miralda-Escud\'{e} \& Gould (2000)]{Miralda00} Miralda-Escud\'{e} J. \& Gould A., 2000, ApJ, 545, 847
\bibitem[Morris (1993)]{Morris93} Morris M., 1993, ApJ, 408, 496
\bibitem[Narayan \& Yi (1995)]{Narayan95} Narayan R. \& Yi I., 1995, ApJ, 444, 231
\bibitem[O'Leary et al. (2009)]{O'Leary09} O'Leary R.M., Kocsis B. \& Loeb A., 2009, MNRAS, 395, 2127
\bibitem[O'Shaugnessey, Gerosa \& Wysocki (2017)]{Oshaugh17} O'Shaughnessey R., Gerosa D. \& Wysocki D., 2017, Phys. Rev. Lett, accepted, arXiv:1704.03879
\bibitem[Paardekooper et al. (2010)]{Paarde10} Paardekooper S.-J., Baruteau C., Crida A. \& Kley W., 2010, MNRAS, 401, 1950
\bibitem[Paczynski \& Witta (1980)]{PW80} Paczynski B. \& Witta P.J., 1980, A\&A, 88, 23 
\bibitem[Paumard et al. (2006)]{Paumard06} Paumard T. et al., 2006, ApJ, 643, 1011
\bibitem[Portegies Zwart et al. (2006)]{PortegiesZ06} Portegies Zwart S.F. et al., 2006, ApJ, 641, 319 
\bibitem[Reines \& Volonteri (2015)]{Reines15} Reines A.E. \& Volonteri M., 2015, ApJ, 813, 82 
\bibitem[Rodriguez et al. (2016)]{Rodriquez16} Rodriquez C. et al. 2016, arXiv. etc. 
\bibitem[Samsing et al. (2014)]{Samsing14} Samsing J., MacLeod M. \& Ramirez-Ruiz E. 2014, ApJ, 784, 71 
\bibitem[S\'{a}nchez-Salcedo \& Chametla (2014)]{Sanchez14} S\'{a}nchez-Salcedo F. J. \& Chametla R.O. 2014, ApJ, 794, 167 
\bibitem[Schawinski et al. (2015)]{Schawinski15} Schawinski K., Koss M., Berney S.  \& Sartori L.F. 2015, ApJ, 451, 2517 
\bibitem[Schnittman \& Buonnano (2007)]{SchnittBuon07} Schnittman J.D. \& Buonanno A. 2007, ApJ, 662, L63 
\bibitem[Sirko \& Goodman (2003)]{Sirko03} Sirko E. \& Goodman J. 2003, MNRAS, 341, 501 
\bibitem[Stahler (2010)]{Stahler10} Stahler S.W., 2010, MNRAS, 402, 1758 
\bibitem[Stone et al. (2017)]{Stone17} Stone N.C. Metzger B.D. \& Haiman Z., 2017, MNRAS, 464, 946 
\bibitem[Thompson et al. (2005)]{Thompson05} Thompson T.A., Quataert E. \& Murray N. 2005, ApJ, 630, 167 


\end{thebibliography}
\end{document}